\documentclass[english,aps,prl,twocolumn,graphics,graphicx
]{revtex4-1}
\usepackage[utf8]{inputenc}

\usepackage{amsmath,mathtools}
\usepackage{amssymb}
\usepackage{graphicx}
\usepackage{verbatim}
\usepackage{braket}
\usepackage{soul}
\usepackage{bm}
\usepackage{gensymb}
\usepackage{diagbox}
\usepackage{siunitx}
\usepackage[T1]{fontenc}
\usepackage[english]{babel}
\usepackage{mathtools}
\usepackage{newtxtext}  
\usepackage{amsmath}
\usepackage{amsfonts}
\usepackage{booktabs}
\usepackage{hyperref}

\renewcommand\vec{\boldsymbol}
\begin{document}
\title{Orbital Magnetization from Uniform and Periodic Magnetic Fields
}
\author{Chunli Huang}
\affiliation{Department of Physics and Astronomy, University of Kentucky, Lexington, Kentucky 40506-0055, USA}
\date{\today} 

\begin{abstract}
Magnetization is thermodynamically defined as the derivative of the
grand potential with respect to a uniform magnetic field. However, a uniform magnetic field makes the kinetic momentum operators
noncommuting and Landau-quantizes the electron motion. This changes
the zero-field momentum-space to Landau-levels and raises a fundamental question: how can the thermodynamic response to a
uniform field be reproduced by a linear-response calculation carried out
in the momentum space of the zero-field problem? We address this question analytically in a quantum Hall ferromagnet
that allows the orbital magnetization $M$ to be computed in a closed form. We first compute $M$ from
the local Hartree--Fock projector response to a periodic magnetic field with
zero net flux. We then compute $M$ from the derivative of the grand potential
with respect to a uniform magnetic field along the St\v{r}eda line. The two
approaches give the same result, even though the first keeps the
Hilbert space fixed while the second changes the Landau-level degeneracy. Their agreement suggests that we should view orbital magnetization as the
energy associated with the spectral flow that gives rise to the St\v{r}eda
formula. Our work provides a tutorial introduction to orbital magnetization and its relation to the St\v{r}eda formula.
\end{abstract}
\maketitle

\section{Introduction}
Orbital magnetization has recently received renewed interest
\cite{vidarte2025real,saati2025theory,chen2026orbital,li2026contribution,
das2024unconventional,das2025microscopic,ye2026quantum,kang2025orbital,
liu2026orbital}, driven in part by the experimental discovery of strongly
correlated phases in moir\'e materials
\cite{sharpe2019emergent,
Zeng2023,Cai2023,Park2023,PhysRevX.13.031037}
and rhombohedral multilayer graphene
\cite{zhou2022isospin,han2023orbital,deng2025superconductivity}
that break time-reversal symmetry through spontaneous valley polarization. At the same time, nano-SQUID
magnetometry \cite{auerbach2025isospin,arp2024intervalley} is now able to measure changes in magnetization
relative to a reference state with high sensitivity, providing new opportunity to understand this fundamental thermodynamic quantity.

The theory of orbital magnetization in solid-state systems was developed
extensively in the mid-2000s, leading to what is called the modern
theory of orbital magnetization
\cite{thonhauser2005orbital,xiao2005berry,shi2007quantum}. For a two-dimensional crystal, the orbital magnetization density is
\begin{equation}
\label{eq:om}
M =
\sum_n \int \frac{d^2 k}{(2\pi)^2}\,
f(\epsilon_{n \vec{k}})
\left[
m_{n}(\vec{k})
+
\frac{e}{\hbar}
\bigl(\mu-\epsilon_{n\vec{k}}\bigr)
\Omega_{n}(\vec{k})
\right],
\end{equation}
where $f(\epsilon_{n \vec{k}})$ is the occupation and  $\mu$ is the chemical potential. Here $m_{n}(\vec{k})$ intrinsic
orbital magnetic moment of a Bloch wave packet (self-rotation) and 
and the $\Omega_{n}(\vec{k})$ is the Berry-curvature contribution.
This formula is remarkable because the thermodynamic definition of
orbital magnetization as the derivative of the energy with respect to
magnetic field \cite{shi2007quantum}, the semiclassical wave-packet
dynamics in phase-space \cite{xiao2005berry}, and the microscopic definition in terms
of the volume average of current loops \cite{thonhauser2005orbital} all
lead to the same expression. Most early discussions, as well as this agreement among definitions, were
developed primarily for noninteracting electrons.

Hartree--Fock approximation (HFA) provides a useful starting point for understanding interaction-induced spin-valley-polarized phases in two-dimensional materials, in part because it contains the exchange energy responsible for the flavor polarization observed in magnetotransport experiments. Recent work
\cite{kang2025orbital,zhu2026orbital,liu2026orbital} has shown that,
within HFA, Eq.~\eqref{eq:om} retains the
same form provided that one uses the quasiparticle eigenstates and
eigenvalues of the Hartree-Fock mean-field Hamiltonian. 
A subtle point clarified in \cite{zhu2026orbital,vignale2026ward} is that while the vector potential couples microscopically to
the bare velocity $\partial_{\vec{k}} T_{\vec{k}}$, where $T_{\vec{k}}$
is the kinetic energy Hamiltonian. Nevertheless, after the required vertex
corrections are included, the velocity entering the
orbital-magnetization formula is the Hartree--Fock velocity
$\partial_{\vec{k}}H_{\vec{k}}^{\rm HF}$, as required by the Ward
identity.

\begin{table}[t]
\centering
\renewcommand{\arraystretch}{1.18}
\setlength{\tabcolsep}{5pt}
\begin{tabular}{lll}
\hline 
& \textbf{Uniform $B$} 
& \textbf{Periodic $\delta B_{\vec q}$} \\
\hline

Vector potential 
& $\displaystyle \vec A(\vec r)=\frac{1}{2}\vec B\times\vec r$
& $\displaystyle \vec A(\vec r)=\vec A_{\vec q}e^{i\vec q\cdot\vec r}$ \\[5pt]

Magnetic flux 
& $\displaystyle \delta\Phi=A\,\delta B\neq 0$
& $\displaystyle \int d^2r \vec B(\vec r)=0$ \\[5pt]

Method
& Landau quantization
& Finite-$\mathbf q$ perturbation \\[5pt]

St\v{r}eda response 
& $\displaystyle 
\frac{1}{A}\frac{\partial N}{\partial B}\bigg|_{\mu}
=\frac{C}{\phi_0}$
& $\displaystyle 
\delta n_{\vec q}
=\frac{C}{\phi_0}\,\delta B_{\vec q}$ \\[8pt]

Orbital magnetization 
& $\displaystyle 
M=-\frac{\partial}{\partial B}(E-\mu N)$
& $\displaystyle 
\delta\Omega_{\vec q}
=-M\,\delta B_{\vec q}$ \\[5pt]

\hline
\end{tabular}
\caption{\label{tab:dictionary}
Comparison between the uniform-field and periodic-field calculations of the St\v{r}eda formula and orbital magnetization. 
A uniform field changes the total magnetic flux, whereas a periodic magnetic perturbation carries zero net flux and therefore preserves the zero-field Hilbert-space. 
We show that the orbital magnetizations obtained from the two approaches agree with each other.
}
\end{table}

The purpose of this manuscript is to resolve some subtleties in defining
orbital magnetization and to provide a deeper understanding of this
fundamental quantity. The problem is the following. A uniform magnetic
field enters quantum mechanics through a vector potential which grows linearly with position, e.g. $\vec{A}=\vec{B}\times\vec{r}/2$.
This field replaces the commuting momentum variables $k_x, k_y$ of the zero-field  Hamiltonian by non-commuting momentum operators $[\Pi_x,\Pi_y]=ieB/\hbar$. Thus, a uniform field changes the kinematic structure of the Hamiltonian and requires Landau quantization. For example, in monolayer graphene, the linear dispersion $\hbar v_Fk$ is reorganized into the well-known Dirac Landau levels, $E_N=\pm v_F\sqrt{2e\hbar NB}$.
The $\sqrt{B}$ dependence cannot be obtained from ordinary perturbation theory around a zero-field momentum eigenstate in the form $\epsilon_n(k)=\epsilon_n^0(k)+B \Sigma_n^{(1)}(k)$. Moreover, changing $B$ also changes the number of flux quanta and hence the Landau-level degeneracy. This raises a basic question: how can orbital magnetization be formulated as a linear response to a uniform magnetic field when a uniform field is not an ordinary perturbation of the zero-field Bloch Hamiltonian, but instead reorganizes the spectrum into Landau levels?

In Ref.~\cite{zhu2026orbital}, following Ref.~\cite{shi2007quantum}, we avoided this difficulty by perturbing the system with a periodic magnetic field carrying zero net flux. The new ingredient in our approach is to study the resulting change of the Hartree-Fock projector. Because the perturbing field carries no net flux, it does not change the total number of flux quanta through the sample, and the response can be computed within a fixed zero-field Hilbert-space using standard perturbation theory. 
Here we show, by an explicit analytical calculation in a simple model, that the two procedures described in Table.~1 give the same orbital magnetization. We then comment on what this equivalence implies.

\section{Model Hamiltonian}
The toy model Hamiltonian we consider is inspired by the adiabatic model
of twisted homobilayers semiconductor \cite{PhysRevLett.132.096602}. The spin-up electron experiences a uniform
effective magnetic field arising from the layer-skyrmion texture, while
the spin-down electron experiences the opposite effective field. The
total system is therefore time-reversal invariant.
The $N$ electron Hamiltonian is
\begin{equation}
    H =
    \sum_{i=1}^{N}
    \frac{
    \left[
    \vec{p}_i
    + e s_i \vec{A}_{\rm sky}(\vec{r}_i)
    + e \vec{A}_{\rm ext}(\vec{r}_i)
    \right]^2
    }{2m}
    +
    \frac{1}{2}
    \sum_{i\neq j}
    V(r_{ij}),
\end{equation}
where $s_i=+1$ for spin up and $s_i=-1$ for spin down. Here the layer-skyrmion vector potential is,
\begin{equation}
    \vec{A}_{\rm sky}(\vec{r})
    =
    \frac{1}{2}\vec{B}_{\rm sky}\times \vec{r},
\end{equation}
where $\vec{B}_{\rm sky}=B_{\rm sky}\hat z$ is uniform and the magnitude depends on the material properties such as twist angles. For simplicity, we do not consider Zeeman coupling.
A uniform external magnetic field $\vec{B}_{\rm ext}$ can be introduced in the conventional way,
\begin{equation}
    \vec{A}_{\rm ext}(\vec{r})
    =
    \frac{1}{2}\vec{B}_{\rm ext}\times \vec{r}.
\end{equation}
Thus the total effective magnetic field seen by particle $i$ is
\begin{equation}
    \vec{B}_i^{\rm eff}
    =
    s_i\vec{B}_{\rm sky}
    +
    \vec{B}_{\rm ext}.
\end{equation}
This model is convenient because the external magnetic field can be
introduced simply by tuning $\vec{B}_{\rm ext}$. 

When $B_{ext}=0$, each spin species forms Landau levels with
degeneracy
\begin{equation}
    N_{\phi}=\frac{B_{\rm sky}A}{\phi_0},
\end{equation}
where $\phi_0=h/e$ is the flux quantum. We study the groundstate at electron density
\begin{equation}
    n_e=\frac{N}{A}=\frac{N_{\phi}}{A}.
\end{equation}
Thus the two spin-projected lowest Landau levels are half filled in
total. We refer to this filling as
$\nu=1$. For a wide range of repulsive interactions \cite{sodemann2024halperin}, including the Coulomb
interaction with weak Landau-level mixing \cite{xu2026localized}, the ground state at $\nu=1$ is the
familiar quantum Hall ferromagnet where electrons completely occupied one spin-projected Landau level,
\begin{equation}
    |\Psi\rangle
    =
    \prod_{X=1}^{N_{\phi}}
    c_{0X\uparrow}^{\dagger}
    |0\rangle ,
\end{equation}
Here $X$ is the guiding-center. For simplicity, we consider the long range Coulomb potential 
\begin{equation}
    V(r_{ij})=\frac{e^2}{\epsilon |r_i-r_j|}
\end{equation}

\section{Periodic-field calculation}

We now review the approach of Ref.~\cite{zhu2026orbital}, where the
orbital magnetization was derived within the mean-field approximation.
Instead of applying a uniform magnetic field, one studies the response of
the Hartree--Fock density matrix to a weak periodic vector potential
\begin{equation}
    \vec{A}(\hat{\vec{r}})
    =
    \vec{A}_{\vec q} e^{i\vec q\cdot \hat{\vec r}} .
\end{equation}
The corresponding magnetic field is
\begin{equation}
    B_{\vec q}
    =
    i\left(\vec q\times \vec{A}_{\vec q}\right)\cdot \hat{z},
\end{equation}
which is also periodic and has zero net flux.  To leading order in
$A_{\vec q}$, the single-particle density matrix changes as
\begin{equation}
    P \rightarrow P+\delta P_{\vec q}.
\end{equation}
where 
\begin{align} \label{eq:delta_Pq}
    \delta P_{\vec q }
    = \frac{e}{\hbar}\sum_{cv\vec k} &
    \Big(\vec{A}_{\vec q}\!\cdot\!\braket{u_{c\vec k}|\nabla_{\!\vec k} u_{v\vec k}}
        \ket{u_{c,\vec k+\vec q}}\bra{u_{v\vec k}}  \nonumber\\
    &+ \vec{A}^*_{-\vec q} \cdot
    \braket{u_{c\vec k}|\nabla_{\!\vec k} u_{v\vec k}}^* 
    \ket{ u_{v \vec k}} \bra{u_{c\vec k-\vec q}} \Big).
\end{align}

Here $v$ and $c$ label the occupied (unoccupied) Hartree--Fock bands.
From $\delta P_{\vec q}$, we found that the induced density has the Streda form,
\begin{equation} \label{eq:streda}
    \delta n_{\vec q}
    =
    \frac{{\rm Tr}\,\delta P_{\vec q}}{A}
    =
    \frac{C}{\phi_0}B_{\vec q},
\end{equation}
where $C$ is the Chern number of the occupied Hartree--Fock band,
\begin{align}
    C=& \label{eq:chern-number}
    \epsilon_{ji}
    \sum_{cv}
    \int\frac{d^2k}{2\pi}\,
    {\rm Im}
    \left[
    \left\langle
    \partial_{k_i}u_{v\vec{k}}
    \middle|
    u_{c\vec{k}}
    \right\rangle
    \left\langle
    u_{c\vec{k}}
    \middle|
    \partial_{k_j}u_{v\vec{k}}
    \right\rangle
    \right] \\
    =&
    \frac{1}{2\pi}
    \int d^2k\,
    \Omega_z(\vec{k}),
\end{align}
For metal, the Chern number $C$ should be replaced by integrated Berry curvature of the occupied states. 

The same density-matrix response that gives rise to the local charge density
also changes the local grand-potential density, thereby defining the orbital
magnetization,
\begin{equation}
    \delta \Omega_{\vec q}
    =
    \frac{{\rm Tr}\,\hat{\Omega}\,\delta P_{\vec q}}{A}
    =
    -M B_{\vec q},
\end{equation}
where $\hat{\Omega}=H^{\rm MF}-\mu \hat{N}$. The resulting expression for
the orbital magnetization is Eq.~(44) of Ref.~\cite{zhu2026orbital},
which we reproduce here:
\begin{align}
    M
    =&
    \frac{e}{\hbar}
    \sum_{cv}
    \int\frac{d^2k}{(2\pi)^2}
    \left(
        \frac{\xi_{c\vec{k}}+\xi_{v\vec{k}}}{2}
    \right) \nonumber\\
    &\times
    \epsilon_{ij}{\rm Im}
    \left[
    \left\langle
    \partial_{k_i}u_{v\vec{k}}
    \middle|
    u_{c\vec{k}}
    \right\rangle
    \left\langle
    u_{c\vec{k}}
    \middle|
    \partial_{k_j}u_{v\vec{k}}
    \right\rangle
    \right] .
    \label{eq:OM-formula-main}
\end{align}
 We have defined
$\xi_{n\vec{k}}=\epsilon_{n\vec{k}}^{\rm HF}-\mu$, where
$\epsilon_{n\vec{k}}^{\rm HF}$ is the Hartree--Fock quasiparticle energy
and $\mu$ is the chemical potential. As shown in the appendix of Ref.~\cite{zhu2026orbital}, Eq.~\eqref{eq:OM-formula-main} can be rewritten in the same form as 
Eq.~\eqref{eq:om}.
Comparing Eq.~\eqref{eq:OM-formula-main} with
Eqs.~\eqref{eq:streda} and \eqref{eq:chern-number}, we see that orbital
magnetization is the energy-weighted response of the same spectral flow
that creates the local density modulation in the Streda formula, with
weight $(\xi_{c\vec{k}}+\xi_{v\vec{k}})/2$.

This perspective provides several useful insights that are not usually
emphasized in the literature. First, the orbital magnetization vanishes
identically in a particle-hole-symmetric two-band model since
$\xi_{c\vec{k}}=-\xi_{v\vec{k}}$. 
Second, the orbital magnetization has no Fermi surface contribution (at zero temperature) because the energy weight vanishes at the Fermi surface. Orbital magnetization is really the response of the occupied states ( Fermi-sea) to external magnetic field. 
This is different from the intrinic anomalous Hall conductivity in metal, where the non-quantized part is a Fermi-surface property \cite{haldane2004berry}. 
Third, in general, computing orbital magnetization requires information about the wave functions and energies of remote unoccupied states. A two-band model is a special case, where the single unoccupied band is fixed once the occupied band is known. Although Eq.~(1) is expressed in terms of sum over
occupied states, once $m_n(\vec k)$ and $\Omega_n(\vec k)$ are known,  the computation of the self-rotation term
\begin{equation}
m_n(\mathbf k)
=
-\frac{ie}{2\hbar}
\left\langle
\nabla_{\mathbf k} u_{n\mathbf k}
\right|
\times
\left[
H_{\mathbf k}-\epsilon_{n\mathbf k}
\right]
\left|
\nabla_{\mathbf k} u_{n\mathbf k}
\right\rangle
\end{equation}
is not determined by the occupied state alone. This is because $H_{\vec k} \left| \nabla_k u_{n\vec k} \right\rangle =\sum_{m}\epsilon_{m\vec k}\langle u_{m\vec k}|\partial_ku_{n\vec k}\rangle$ and $m$ includes unoccupied states too.
In fact, only the interband Berry connection between occupied and unoccupied states enters the orbital magnetization formula, as shown in Eq.~\eqref{eq:OM-formula-main}. Fundamentally, this is because when we evaluate the expectation value of the current-loop operator
\(\langle n| \mathbf r\times \mathbf v |n\rangle\), one operator (e.g.~$v$) can take the occupied state to the unoccupied states
 and then the other operator (e.g.~$r$) bring it back.

Let us now apply this formalism to the $\nu=1$ quantum Hall ferromagnet
$|\Psi\rangle$ defined in previous section. We set $B_{\rm ext}=0$ and apply a periodic magnetic
field. This field induces both a density modulation,
$\delta n_{\vec q}=C\delta B_{\vec q}/\phi_0$, and a change in the
grand-potential density,
$\delta \Omega_{\vec q}=-M\delta B_{\vec q}$. Then, we compute $M$ using
Eq.~\eqref{eq:OM-formula-main}. Since Eq.~\eqref{eq:OM-formula-main} is written in momentum space, we
first can carry out a basis transformation from guiding-center to
magnetic Bloch states $|u_{n\vec{k}}\rangle$. We choose the real-space
magnetic unit cell to enclose one flux quantum, so that the number of
allowed $\vec{k}$ points in the magnetic Brillouin zone is $N_{\phi}$. The shape of the magnetic Brillouin zone can be arbitrary. The quantum Hall
ferromagnet can then be written equivalently as
\begin{equation}
    |\Psi\rangle
    =
    \prod_{\vec{k}}^{N_\phi}
    c_{0\vec{k}\uparrow}^{\dagger}
    |0\rangle .
\end{equation}
The area of the magnetic Brillouin zone  $A_{\rm MBZ}$ is such that,
\begin{equation}
    \frac{A_{\rm MBZ}}{4\pi^2}
    =
    \frac{1}{2\pi\ell_B^2}\;,\;\ell_B=\sqrt{\frac{\hbar}{eB_{sky}}}
\end{equation}
We emphasize that because the perturbing magnetic field $B_{\vec q}$
carries zero net flux, this set of magnetic Bloch states remains fixed
throughout the calculation. 

The Hartree--Fock quasiparticle energies are ``flat-bands'':
\begin{equation}
        \xi_{n\bm k}
    =
    \epsilon_{n}^{\rm HF}-\mu
    =
    \left(n+\frac{1}{2}\right)\hbar\omega_c
    +\Sigma_n-\mu,
\end{equation}
where $n=0,1,2...$ and 
\begin{equation}
    \Sigma_n =-\sum_{X'}\langle0X,nX'|V|nX',0X\rangle,
\end{equation}
is the Fock self-energy of an
electron in the $n$-th same-spin Landau level due to the filled
$n=0$ Landau level.

In the present problem, the occupied band is
$v=(n=0,\uparrow)$, and the unoccupied bands are the higher same-spin
Landau levels, $c=(n,\uparrow)$ with $n=1,2,3,\ldots$ since the applied vector
potential couples does not flip spin.
Because the Landau-level does not disperse with $\vec k$, we can take 
the energy factor in
Eq.~\eqref{eq:OM-formula-main}
outside the momentum integral and obtain
\begin{align}
    M
    =
    \frac{e}{\hbar}
    \sum_{n=1}^{\infty}
    \left[
        \frac{n+1}{2}\hbar\omega_c
        +
        \frac{\Sigma_n+\Sigma_0}{2}
        -
        \mu
    \right]
    \mathcal J_n .
    \label{eq:Mz-before-Jn}
\end{align}
The term in the square bracket is the average energy between $n$-th unoccupied Landau level and the $n=0$ occupied Landau level. The quantity $\mathcal J_n$ is the n-th Landau level contribution to the
Berry-curvature integral,
\begin{align}
    \mathcal J_n
    =
    \int_{\rm MBZ}
    \frac{d^2k}{4\pi^2}
    \epsilon_{ij}
    \Im\left[
    \braket{\partial_{k_i}u_{0\bm k}|u_{n\bm k}}
    \braket{u_{n\bm k}|\partial_{k_j}u_{0\bm k}}
    \right] .
    \label{eq:Jn-def}
\end{align}
Because the mean-field approximation for a parabolic-band electron gas
does not change the Landau-level wavefunctions, the magnetic Bloch
wavefunctions have the same form as in the noninteracting problem.  The
reason is because the quantum Hall ferromagnet is an isotropic electron-fluid.  Therefore, we can use the Feynman--Hellmann relation
and Landau-level ladder operators to evaluate the matrix elements. Since
the kinetic momentum $\vec{\Pi}$ is linear in the ladder operators
$a$ and $a^\dagger$, the $n=0$ Landau level only mixes with the
$n=1$ Landau level:
\begin{align}
    \braket{u_{n\vec k}|\partial_{k_x}u_{0\vec k}}
    &=
    -\frac{\ell_B}{\sqrt{2}}\delta_{n,1},
    \\
   \braket{u_{n\vec k}|\partial_{k_y}u_{0\vec k}}
    &=
    +\frac{i\ell_B}{\sqrt{2}}\delta_{n,1}.
\end{align}
\begin{align}
    \epsilon_{ij}
    \Im\left[
    \braket{\partial_{k_i}u_{0\bm k}|u_{n\bm k}}
    \braket{u_{n\bm k}|\partial_{k_j}u_{0\bm k}}
    \right]
    =
    -\ell_B^2\delta_{n,1}.
\end{align}
It follows that
\begin{align}
    \mathcal J_n=
    -\ell_B^2\delta_{n,1}
    \frac{A_{\rm MBZ}}{4\pi^2}=
    -\frac{1}{2\pi}\delta_{n,1}. \label{eq:Jn-result}
\end{align}
Substituting Eq.~\eqref{eq:Jn-result} into
Eq.~\eqref{eq:Mz-before-Jn}, only the $n=1$ unoccupied Landau level
contributes:
\begin{align}
    M_z
    &=
    -\frac{e}{2\pi\hbar}
    \left[
        \hbar\omega_c
        +
        \frac{\Sigma_1+\Sigma_0}{2}
        -
        \mu
    \right] \nonumber \\
    &=
    \frac{1}{\phi_0}
    \left[
        \mu
        -
        \hbar\omega_c
        +
        \frac{3}{4}
        \sqrt{\frac{\pi}{2}}
        \frac{e^2}{\epsilon\ell_B}
    \right].
    \label{eq:Mz-final-explicit}
\end{align}
We used $\Sigma_1=\frac{1}{2}\Sigma_0$ in the last equation. This relation can be derived from standard gaussian integral,
\begin{align}
    \Sigma_n
    =
    -\int\frac{d^2q}{(2\pi)^2}\,
    V(q)\,
    |F_{n0}(\bm q)|^2,
\end{align}
where the form factors are
\begin{align}
    |F_{00}(\bm q)|^2
    =
    e^{-q^2\ell_B^2/2},\quad
    |F_{10}(\bm q)|^2
    =
    \frac{q^2\ell_B^2}{2}
    e^{-q^2\ell_B^2/2}.
\end{align}
Therefore
\begin{align}
    \Sigma_0
    &=
    -\frac{e^2}{\epsilon}
    \int_0^\infty dq\,
    e^{-q^2\ell_B^2/2}
    =
    -\sqrt{\frac{\pi}{2}}
    \frac{e^2}{\epsilon\ell_B},
    \\
    \Sigma_1
    &=
    -\frac{e^2}{\epsilon}
    \int_0^\infty dq\,
    \frac{q^2\ell_B^2}{2}
    e^{-q^2\ell_B^2/2}
    =
    -\frac{1}{2}
    \sqrt{\frac{\pi}{2}}
    \frac{e^2}{\epsilon\ell_B}
    \label{eq:Sigma1-half-Sigma0}
\end{align}

\section{Uniform-field calculation}

As mentioned in Section II,  adding a
uniform external magnetic field in our toy model simply changes the effective magnetic
field seen by each spin species,
\begin{equation}
    \vec{B}_i^{\rm eff}
    =
    s_i\vec{B}_{\rm sky}
    +
    \vec{B}_{\rm ext}.
\end{equation}
For the $\nu=1$ quantum Hall ferromagnet considered here where only spin up is occupied, they experience a net magnetic field $    \vec{B}_{\uparrow}^{\rm eff}
    =
    \vec{B}_{\rm sky}
    +
    \vec{B}_{\rm ext}.$

We consider the response at fixed chemical potential, with $\mu$ lying
inside the exchange gap. As the external flux is increased, the degeneracy
of the spin-up Landau level increases. Along the St\v{r}eda line, $\left.\partial n/\partial B\right|_{\mu}=1/\phi_0$,
the particle number changes in such a way that the lowest Landau level
remains completely filled. Thus the ground
state is a family of quantum Hall ferromagnets parametrized by the
external field,
\begin{equation}
    |\Psi(\lambda)\rangle
    =
    \prod_{X=1}^{N(\lambda)}
    c_{0X\uparrow}^{\dagger}|0\rangle .
\end{equation}
Here $N(\lambda)$ is the Landau-level degeneracy in the presence of the
external magnetic field,
\begin{equation}
    N(\lambda)
    =
    \frac{(B_{\rm sky}+B)A}{\phi_0}
    =
    N_{\phi}(1+\lambda),
    \qquad
    \lambda=\frac{B}{B_{\rm sky}} .
\end{equation}
This is what we mean when we say that the uniform-field calculation takes
place in a varying magnetic Hilbert space. In a crystal, this is
more dramatic. When the flux per unit cell is rational, $\alpha=\frac{\Phi_{\rm cell}}{\phi_0}=\frac{p}{q}$, an ordinary Bloch band splits into $q$ magnetic subbands. For example, at
$\alpha=1/5$, one ordinary band splits into five magnetic Bloch bands. If
the flux is decreased slightly to $\alpha=9/50$, each of the magnetic
Bloch bands at $\alpha=1/5$ is further split into ten narrower magnetic
subbands. The continuity of this spectrum between rational $\alpha$ and irrational $\alpha$ is discussed in
Ref.~\cite{hofstadter1976energy}.

The stability of $|\Psi(\lambda)\rangle$ as a function of $\lambda$ was
studied in detail in Ref.~\cite{xu2026two}. For parameters relevant to
twisted MoTe$_2$, the incompressible state along the St\v{r}eda line
pointing away from charge neutrality remains stable. We can therefore
compute the orbital magnetization by differentiating the grand potential
with respect to the external magnetic field at fixed chemical potential,
\begin{equation}
    M
    =
    -\frac{1}{A}
    \left.
    \frac{\partial \Omega}{\partial B}
    \right|_{\mu}.
\end{equation}
The expectation value of the Hamiltonian in the state
$|\Psi(\lambda)\rangle$ is
\begin{align}
    E(\lambda)
    =& \langle \Psi(\lambda)|H|\Psi(\lambda)\rangle\\
    =&
    N(\lambda)
    \left[
        \frac{1}{2}\hbar\omega_c(\lambda)
        +
        \frac{1}{2}\Sigma_0(\lambda)
    \right],
\end{align}
 where the first term is the zero-point energy and the second term is the exchange energy $    \Sigma_0(\lambda)
    =
    -\sqrt{\frac{\pi}{2}}
    \frac{e^2}{\epsilon \ell(\lambda)}$. The cyclotron frequency and magnetic length depends on $\lambda$,
\begin{align}
    \omega_c(\lambda)
    &=
    \frac{e(B_{\rm sky}+B)}{m}
    =
    \omega_c(1+\lambda) \\
    \ell(\lambda)
    &=
    \sqrt{\frac{\hbar}{e(B_{\rm sky}+B)}}
    =
    \frac{\ell_B}{\sqrt{1+\lambda}},
\end{align}
Here $\omega_c=eB_{\rm sky}/m$ and
$\ell_B=\sqrt{\hbar/(eB_{\rm sky})}$ are the values at zero external field. The grand potential changes with $\lambda$ as
\begin{equation}
    \Omega(\lambda)= E(\lambda)-\mu N(\lambda)
\end{equation}
Explicitly, the grand-potential density is
\begin{align}
    \frac{\Omega(\lambda)}{A}
    =
    \frac{B_{\rm sky}}{\phi_0}
    \bigg[&
        \frac{1}{2}\hbar\omega_c(1+\lambda)^2
        -
        \frac{1}{2}\sqrt{\frac{\pi}{2}}
        \frac{e^2}{\epsilon \ell_B}(1+\lambda)^{3/2}
        \nonumber \\&-
        \mu(1+\lambda)
    \bigg].
\end{align}
This expression is valid for all external field.
 The orbital magnetization is then obtained by evaluating the slope of the grand
potential at $\lambda=0$,
\begin{align}
    M
    &=
    -\left.
    \frac{\partial}{\partial B}
    \frac{\Omega}{A}
    \right|_{\mu,B=0}  \nonumber \\
    &=
    -\frac{1}{B_{\rm sky}}
    \left.
    \frac{\partial}{\partial \lambda}
    \frac{\Omega(\lambda)}{A}
    \right|_{\mu,\lambda=0}  \nonumber \\
    &=
    -\frac{1}{\phi_0}
    \left.
    \left[
        \hbar\omega_c(1+\lambda)
        -
        \frac{3}{4}
        \sqrt{\frac{\pi}{2}}
        \frac{e^2}{\epsilon\ell_B}
        (1+\lambda)^{1/2}
        -
        \mu
    \right]
    \right|_{\lambda=0} \nonumber \\
    &=
    \frac{1}{\phi_0}
    \left[
        \mu
        -
        \hbar\omega_c
        +
        \frac{3}{4}
        \sqrt{\frac{\pi}{2}}
        \frac{e^2}{\epsilon\ell_B}
    \right].
\end{align}
The orbital magnetization obtained from the ground-state
energy calculation in a uniform field agrees with the periodic-field result in the previous section Eq.~\eqref{eq:Mz-final-explicit}. The agreement is nontrivial, in my view. For example, the factor of $3/4$ in front of the self-energy $\sqrt{\frac{\pi}{2}}\frac{e^2}{\epsilon\ell_B}$ appears in two completely different ways.

\section{Discussion}
The formula derived in Ref.~\cite{zhu2026orbital} perturbs the zero-field Hartre-Fock density matrix with a periodic magnetic field. To leading order, the perturbation induces
particle-hole components in the density matrix, mixing occupied states of the
Hartree-Fock Slater determinant with unoccupied states. This
occupied--unoccupied mixing is the microscopic origin of the local
St\v{r}eda response. Regions with \(\delta B(\mathbf r)>0\) has an
enhanced local density, while regions with \(\delta B(\mathbf r)<0\) has a
reduced local density. For parabolic-dispersion electron gas, this is
generated by mixing the occupied
$n=0$ Landau level with the unoccupied $n=1$ Landau level. 
The orbital magnetization is equivalent to the energy
associated with this field-induced mixing of the wavefunction \cite{zhu2026orbital}. The energy-weight factor is given by the
average Hartree-Fock energy of the occupied $n=0$ and unoccupuied $n=1$ level. The exchange part of this average energy is $
    \frac{\Sigma_0+\Sigma_1}{2}=
    \frac{3}{4}\Sigma_0$.

By contrast, in the uniform-field calculation, increasing $B$
increases the Landau-level degeneracy. At fixed chemical potential inside
the exchange gap, this increases the number of electrons in the
majority-spin lowest Landau level. The information about the unoccupied Landau levels do not
enter explicitly in this method. Instead, the same factor of $3/4$
arises from differentiating the total exchange-energy density with respect to $B$. 

The agreement between the two calculations is therefore nontrivial. 
In the periodic-field calculation, the Hilbert space is held fixed, and the
magnetic perturbation induces Landau-level mixing between the $n=0$ and $n=1$ Landau level. In the uniform-field calculation, by contrast, we evaluate the
groundstate energy in the Hilbert space appropriate to the new value of $B$,
where the Landau-level degeneracy has changed. Their agreement shows that
the local projector response \(\delta P_{\mathbf q}\) encodes the same physics
as the global change in Landau-level degeneracy.

The spirit of linear response is to compute things without re-diagonalizing the Hamiltonian but nevertheless uniform B-field described by unbounded vector potential necessarily involves Landau-quantization of the zero field Hamiltonian. 
However, if we understand orbital magnetization as the energy associated with spectral flow, one is global spectral flow, the other is local spectral flow, then it is not so surprising that they give rise to the same result. The microscopic details of spectral flow, say mixing 0 and 1 Landau level, are system dependent, but the resulting orbital magnetization is the same.

A similar comparison was
made in Ref.~\cite{kang2025orbital} for the Rashba model.
Since the free energy is known for arbitrary external field in this model, it can also be used to test the magnetic susceptibility by computing $\partial^2\Omega/\partial \lambda^2$ from the
uniform-field free energy and compare it with a zero-field momentum space calculation.

\section{St\v{r}eda formula and the quantum Hall effect}

We have presented orbital magentization as an energy associated with changing particle via Streda formula. It is therefore natural to ask whether the same viewpoint can be related to the quantum Hall effect. It turns out that it can, provided the
bulk charge response is gapped so that the long-wavelength and
low-frequency limits commute.

Let us assume, for the moment, that the St\v{r}eda formula in
Eq.~\eqref{eq:streda} remains valid at finite frequency,
\begin{equation}
    \delta n(\vec q,\omega)
    =
    \frac{C}{\phi_0}
    \delta B_z(\vec q,\omega).
\end{equation}
We take $\vec q=q\hat{x}$ and choose a transverse vector potential
\begin{equation}
    \vec A(\vec r,t)
    =
    A_y(\vec q,\omega)
    e^{i(qx-\omega t)}
    \hat y ,
\end{equation}
with vanishing scalar potential. Then,
\begin{align}
    \delta B_z(\vec q,\omega)
    =
    i q A_y(\vec q,\omega),
\,
    E_y(\vec q,\omega)
    =
    i\omega A_y(\vec q,\omega).
\end{align}
Therefore the ``finite-frequency'' St\v{r}eda formula relates particle density to electric field,
\begin{equation}
    \delta n(\vec q,\omega)
    =
    \frac{C}{\phi_0}
    i q A_y(\vec q,\omega)
    =
    \frac{C}{\phi_0}
    \frac{q}{\omega}
    E_y(\vec q,\omega).
\end{equation}
We can next use the continuity equation for particle number,
\begin{equation}
    -i\omega\,\delta n(\vec q,\omega)
    +
    i q\, j_x(\vec q,\omega)
    =
    0,
\end{equation}
to arrive at the quantized Hall response, 
\begin{equation}
    j_x(\vec q,\omega)
    =
    \frac{\omega}{q}
    \delta n(\vec q,\omega)
    =
    \frac{C}{\phi_0}
    E_y(\vec q,\omega).
\end{equation}
Multiplying by the electron charge gives the quantized Hall conductivity,
\begin{equation}
    \sigma_{xy}
    =
    C\frac{e^2}{h}.
\end{equation}

There is  a very important assumption hidden in this derivation. The
St\v{r}eda formula is a thermodynamic response: it is derived at
$\omega=0$ and then taking the long-wavelength limit $q\rightarrow0$. The Hall conductivity is a transport response: we first
take the uniform limit $q\rightarrow0$ and then
$\omega\rightarrow0$ later. The derivation above is valid only if these two limits
commute. These two limits are expected to commute for many-body states with no gapless bulk
charge excitations. In that case the electromagnetic response is regular near
$(q,\omega)=(0,0)$, and the same coefficient controls both the static
density response and the Hall current.
In a metal, these two limits clearly do not commute due to the Fermi surface \cite{giuliani2008quantum}, and the Hall conductivity derived from our argument above is not the full Hall conductivity. In fact, our
argument can only give the $\sigma_{xy}^{II}$ contribution in
Ref.~\cite{streda1982theory}. A proper calculation for a metal must also
recover the Fermi-surface contribution, $\sigma_{xy}^{I}$ in Ref.~\cite{streda1982theory}.
\newline

\textbf{Acknowledgment} I am grateful to Giovanni Vignale for helpful discussions and for encouraging me to publish this note. This work is supported by the U.S. Department of Energy, Office of Science, Office of Basic Energy Sciences, under Award Number DE-SC-0024346.

\bibliographystyle{ieeetr}

\bibliography{reference_OM}

\appendix

\pagebreak
\newpage

\appendix

\end{document}